\documentclass[twocolumn,numbers,amsmath]{revtex4}
\usepackage{graphicx}
\usepackage{dcolumn}
\usepackage{bm}
\usepackage{epsfig}

\usepackage{wrapfig}

\begin{document}

\title{Static corrections versus dynamic correlation effects  \\ 
in the valence band Compton profile spectra of Ni}    
\author{L. Chioncel$^{a,b}$, D. Benea$^{c,d}$, S. Mankovsky$^{d}$, H. Ebert$^{d}$, and J. Min\'ar$^{d,e}$}
\affiliation{$^{a}$ Theoretical Physics III, Center for Electronic
Correlations and Magnetism, Institute of Physics, University of
Augsburg, D-86135 Augsburg, Germany}
\affiliation{$^b$ Augsburg Center for Innovative Technologies, University of Augsburg, 
D-86135 Augsburg, Germany}
\affiliation{$^{c}$Faculty of Physics, Babes-Bolyai University,
Kog\u alniceanustr 1, Ro-400084 Cluj-Napoca, Romania} 
\affiliation{$^{d}$Chemistry Department, University Munich,
Butenandstr.~5-13, D-81377 M\"unchen, Germany} 
\affiliation{New Technologies - Research Center, University of West Bohemia, Univerzitni 8, 306 14 Pilsen, Czech Republic}

\begin{abstract}
We compute the Compton profile of Ni using the Local Density Approximation of 
Density Functional Theory supplemented with electronic correlations treated
at different levels. The total/magnetic Compton profiles show not only 
quantitative but also qualitative significant differences depending weather 
Hubbard corrections are treated at a mean field +U or in a more sophisticated 
dynamic way. Our aim is to discuss the range and capability of electronic 
correlations to modify the kinetic energy along specific spatial directions.
 The second and the fourth order moments of the difference in the Compton 
profiles are discussed as a function of the strength of local Coulomb 
interaction $U$.  
\end{abstract}


\maketitle

\section{Introduction}

The study of physical and chemical properties of transition metals is still an 
extremely active experimental field and at the same time is the subject of
extensive theoretical studies. The fascinating aspect of $d$-electron 
systems is the possible interplay of relativistic and electron 
correlation effects that has long been questioned. 
Ab-initio methods provide the framework in 
which relativity and correlations may be treated at equal footing. One 
notable example is the magnetic anisotropy energy for Ni. Experimentally it 
is known that the easy axis is along the [111] direction and the energy cost 
to rotate the magnetic moment axis into [001] direction is about $3~\mu$eV 
per atom\cite{AM77}. LSDA+U calculations~\cite{YS01} accounting for 
spin-orbit and non-collinear coupling have been employed and showed to 
reproduce these values for relatively small value of the local Coulomb 
interaction $U = 1.9$~eV and $J = 1.2$~eV. Changes in the topology of the 
Fermi surface were discussed in the context of magneto-crystalline anisotropy 
of Ni~\cite{G78}. 
These changes were recently addressed using the Gutzwiller variational 
theory with ab-initio parameters which showed the importance of the spin-orbit coupling~\cite{BG08}. 

As a matter of fact, nickel is perhaps the most studied electronic system. 
In the ordered ferromagnetic phase the vast majority of band structure 
calculations within the Local Density Approximation to the Density Functional 
Theory (DFT) converge to a value for the magnetic moment of $\approx 0.62~\mu_B$, 
which is a slightly overestimation of the experimental data. The orbital 
contribution amounts up to $10\%$ and the 
spin moment is found to be around $0.56~\mu_B$. The Generalized Gradient 
Approximation (GGA) add gradient correction to the local density approximation, 
does not change upon the value of the magnetic moment, however improve on 
the equilibrium lattice parameter and bulk modulus. The exchange splitting 
in both LSDA/GGA is in the range of $0.7$ to $0.75$~eV~\cite{LC91}, while 
experimental data 
are situated between $0.3-0.5$~eV~\cite{EH78,DG78,HK79,EH80,EP80}. The valence 
band photoemission spectra of Ni shows a $3d$-band width that is about $30 \%$ 
narrower than the value obtained from the LSDA calculations. It is known that 
LSDA cannot reproduce the dispersionless feature at about 6 eV binding energy, 
the so-called 6 eV satellite \cite{GBP+77}. An 
improved description of correlation effects for the $3d$ electrons via the 
combined Local Density Approximation and Dynamical Mean Field 
Theory~\cite{MV89,GK96,KV04},
LSDA+DMFT, gives the width of the occupied $3d$ bands of Ni properly, reproduce 
the exchange splitting and the 6 eV satellite structure in the valence 
band~\cite{LK01,CVA+03,MCP+05,BME+06,GM07,SFB+09,GM12,SBM+12}.

Momentum space quantities such as the spin-dependent electron momentum 
density distribution have been calculated using various 
methods~\cite{KA90,DDG+98} mostly employing the LSDA. 
In addition to that, magnetic Compton scattering can provide a sensitive method 
of investigating the spin-dependent properties. For example, in Ni, it has 
already been shown that the negative polarization of the $s$- and $p$-like band 
electrons can be observed \cite{KA90,TB90}. Although the total spin moment 
is well reproduced by theory, the degree of negative polarization at low 
momentum, where these electrons contribute, is typically underestimated. 
This discrepancy is often regarded as being due to the insufficient 
treatment of correlation present in the LSDA exchange-correlation 
functional at low momentum~\cite{KA90}. 
Early studies of electronic correlations in band structure 
calculations for the Compton profiles in Li and Na (alkali metals) 
have been performed by Eisenberger et al.~\cite{ei.la.72} and by 
Lundqvist and Lynden~\cite{lu.ly.71}. In the former study,
the linear response theory to the atomic potential in the random phase
approximation is used~\cite{ei.la.72}, while in the later the 
orthogonalized plane wave method for the homogeneous interacting 
electron gas data~\cite{lu.ly.71} has been employed. Although both 
studies have been succesful to describe the momentum densities and 
the Compton profiles they are not suitable for transition-metal systems. 
Later on in the study of transition metals 
Bauer et al.~\cite{BS83,B84,BS84,BS85} investigated extensively 
the role of local and non-local DFT functionals for the problem 
of electron-electron correlation effects pointing out several inconsistencies 
and improving the agreement of theoretical difference profiles with the 
experimental data. We have studied recently whitin the framework of 
LSDA+DMFT the directional Compton profile $J({\bf p}_z)$ for both Fe and 
Ni~\cite{BMC+12,CBE+14}. 
The second moment of the Compton profile difference allowed to quantify
the momentum space anisotropy of the electronic correlations of Fe and Ni.
The changes in the shape and magnitude of the anisotropy have been discussed
as a function of the strength of the Coulomb interaction $U$. According to our
results Ni has a larger momentum space anisotropy of the second moment of
the total Compton profile in comparison with Fe~\cite{CBE+14}. 

The aim of this paper is twofold. First, we perform a comparison of the 
computed magnetic Compton profiles at a mean-field (LSDA+U) level and 
beyond within 
the framework of dynamical mean field theory (LSDA+DMFT). Secondly, we 
extent our previous work on computing moments of directional Compton 
profiles~\cite{CBE+14} and analyze corrections to the 
kinetic energy. In particular, we compute the magnitude of the fourth 
moment that is 
proportional to relativistic kinetic energy corrections that arises from 
the variation of electron mass with velocity. We discuss therefore  
the extend to which electronic correlations can influence the relativistic 
correction to the kinetic energy.

In the following subsections we analyze the magnetic Compton 
(Sec. \ref{sec:mcp}), the total Compton and the difference profiles 
(Sec. \ref{sec:cp_diff}). Subsection Sec. \ref{sec:ke} analyses the 
effects of electronic correlations upon the kinetic energy of bonded 
electrons. We conclude the present paper in section Sec. \ref{sec:conc}.

\section{Magnetic and Total Compton profiles in the presence of electronic correlations}
\label{sec:tf}
We performed the electronic structure calculations using the spin-polarized 
relativistic Korringa-Kohn-Rostoker (SPR-KKR) method in the atomic sphere 
approximation (ASA) \cite{EKM11}. The exchange-correlation potentials 
parameterized by Vosko, Wilk and Nusair \cite{VWN80} were used for the LSDA 
calculations. For integration over the Brillouin zone the special points
method has been used \cite{MP76}. Additional calculations have been performed 
with the many-body effects described by means of dynamical mean field theory 
(DMFT)~\cite{MV89,GK96,KV04} using the relativistic version of the so-called 
Spin-Polarized T-Matrix Fluctuation Exchange approximation \cite{KL02, PKL05} 
impurity solver. In this case a charge and self-energy self-consistent 
LSDA+DMFT scheme for correlated systems based on the KKR 
approach~\cite{MCP+05,MM09,MI11} has been used. The realistic multi-orbital
interaction has been parameterized by the average screened Coulomb 
interaction $U$ and the Hund exchange interaction $J$. Despite the recent 
developments allow to compute the dynamic electron-electron interaction 
matrix elements exactly \cite{AIG+04}, we consider in the present work
the values of $U$ and $J$ as parameters for the sake of convenience of our
discussions. It was shown that the static limit of the screened-energy 
dependent Coulomb interaction leads to a U parameter in the energy range of 
1 and 3 eV for all $3d$ transition metals. As the $J$ parameter is not 
affected by screening it can be calculated directly within the LSDA and
is approximately the same for all 3$d$ elements, i.e $J \approx 0.9$~eV. 
In our calculations we used values for the Coulomb parameter in the range 
of $U = 2.0$ to 3.0 eV and the Hund exchange-interaction $J = 0.9$~eV.

The KKR Green function formalism allows to compute Compton profiles 
$J_{\bf K}(p_z)$ and magnetic Compton profiles $J_{mag,\bf K}(p_z)$
(MCPs) in a straightforward way~\cite{SGST84,BME06,DB04}. In the case 
of a magnetic sample, the spin resolved momentum densities are computed
 within the framework of LSDA and LSDA+DMFT
approaches using the Green's functions in momentum space, as follows:  
\begin{equation}\label{e7}
n_{m_s}(\vec p)={-\frac{1}{\pi} \Im  \int_{-\infty}^{E_F}
G_{m_s}^{LSDA(+DMFT)}(\vec p,\vec p,E)dE}\,\nonumber.
\end{equation}
where $m_s=\uparrow(\downarrow)$.
%
%
The total electron ($n_{\uparrow}(\vec p) + n_{\downarrow}(\vec p)$) and
spin ($n_{\uparrow}(\vec p) - n_{\downarrow}(\vec p)$) momentum densities
projected onto the direction ${\bf K}$ defined by the scattering vector, 
allows to define the (Magnetic) Compton profile as a double integral in 
the momentum plane perpendicular to the scattering momentum 
$\vec p_z  ||$ {\bf K}:
\begin{eqnarray}\label{mcs3}
J_{\bf K}^{LSDA(+DMFT)}(p_z)=
\int \int [ n_{\uparrow}(\vec p) + n_{\downarrow}(\vec p) ] dp_x dp_y \nonumber  \\
J_{mag,\bf K}^{LSDA(+DMFT)}(p_z)=
\int \int [ n_{\uparrow}(\vec p) - n_{\downarrow}(\vec p) ] dp_x dp_y \nonumber.
\end{eqnarray}
A useful quantity in our analysis is the difference of Compton profiles taken along the same momentum space
direction with or without including electronic correlations:
\begin{equation}\label{mcs_dmft-lsda}
\Delta J_{\bf K}(p_z) = J_{\bf K}^{+U/DMFT}(p_z) - J_{\bf K}^{LSDA}(p_z).
\end{equation}
In our further analysis the anisotropies of the Compton profile
\begin{equation}\label{J_kkp}
\Delta J_{\bf K, K^\prime}(p) = J_{\bf K}(p_z) - J_{\bf K^\prime}(p_z)
\end{equation}
are also studied using different local exchange-correlation potentials: 
the ``pure'' LSDA, and the supplemented LSDA+U and LSDA+DMFT ones. 
The electron momentum densities are usually calculated for the
principal directions ${\bf K}=[001], [110]$ and $[111]$ using an 
rectangular grid of 200 points in each direction. The maximum value of 
the momentum in each 
direction is 8 a.u.. The resultant Magnetic-Compton (Compton) profiles were 
normalized to their respective calculated magnetic spin moments (number
of valence electrons). 

\subsection{Magnetic Compton profiles}
\label{sec:mcp}
\begin{figure}[h]
   \includegraphics[width=0.99\linewidth, clip=true]{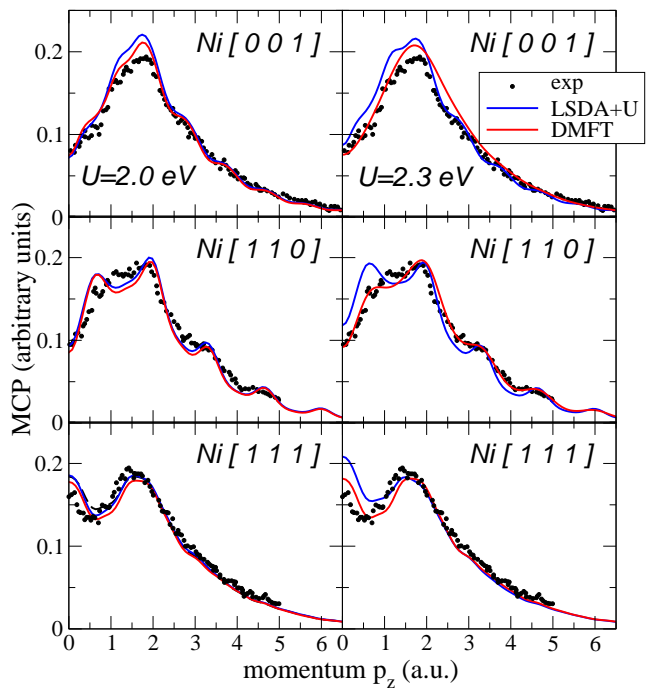}
\caption {\label{Fig:figure1}(color on-line) Compton profiles of Ni
along the principal directions $[001]$, $[110]$ and $[111]$. 
Left column compares the LSDA+U/DMFT for $U = 2.0$~eV with the results 
for $U = 2.3$~eV. For both calculations $J = 0.9$~eV and $T = 400$~K have 
been used. The computed profiles LSDA (black-dotted), LSDA+U (blue-dashed) 
and LSDA+DMFT (red dashed-dot) are plotted in comparison to the experimental 
spectra (black-dotted). The calculated MCP profiles were convoluted with 
the experimental resolution 0.2 a.u.. Data were taken from Dixon et. al.
\cite{DDG+98}. }
\end{figure}

The computed magnetic Compton profiles are shown in Fig. \ref{Fig:figure1} 
along the principal directions [001], [110] and [111]. The profiles seen 
in the left/right columns of Fig.\ \ref{Fig:figure1} have been obtained 
using the values of $U = 2/2.3$~eV for the Coulomb and $J = 0.9$~eV for 
the exchange parameters and the temperature of 400~K. The calculated spin 
moment for LSDA is  $0.61 \mu_B$ while both LSDA(+U/DMFT) results give 
$0.59 \mu_B$. The theoretical MCPs have been convoluted 
with experimental momentum resolution ~\cite{DDG+98} and all the areas
are normalized at the coresponding spin moments. 
Our LSDA results~\cite{CBE+14} are in agreement with the previous 
published results, obtained with LMTO~\cite{DDG+98} or the 
FLAPW-LSDA~\cite{KA90,BZK00}.

The most obvious feature for all principal directions is a significant 
discrepancy between experiment and theory for $p_z < 2$~a.u. 
(see Fig. \ref{Fig:figure1}).
We notice the large dips in the [110] and [111] profiles near $p_z = 0$ a.u., 
which were ascribed partially to the $s$- and $p$-like electrons, 
but also to a pronounced drop in the contribution from the fifth
band~\cite{DDG+98}. 
%

The results of computations for  the average Coulomb parameter 
$U = 2.0$~eV are shown on the  left column of Fig.\ \ref{Fig:figure1}. 
Along the [001] direction and around $p_z = 0$~a.u. all LSDA(+U/DMFT) 
results seem to get close to the experimental data. Most significant 
differences are in the momentum range of $1$~a.u.$< p_z < 2$~a.u., 
where also the maximum of the profile is located. Along the $[110]$
direction, both LSDA+U and LSDA+DMFT give similar results, 
overestimating the first maximum at around $0.5$~eV, show a minimum at
around $1$~eV, instead of a maximum seen in the experiment and underestimate
the experimental results in several regions above $2$~a.u. Along the
$[111]$ direction, the maximum at $p_z = 0$~a.u. is overestimated by all
computations: the slight improvement of DMFT is not really significant, LSDA+U
get very close to the maximum at around $2$~a.u. Overall dynamic
correlations do really not improve significantly the agreement with the
experimental data, as already at the level of LSDA reasonable good
results has been obtained.  

For a slightly larger value of $U = 2.3$~eV the LSDA+U results start to depart 
more from the experimental data, while on contrary the DMFT results improve 
the agreement significantly. Along the $[001]$ in the entire low momentum 
region $p_z < 2$~a.u. LSDA+U overestimates the spectrum, however for larger 
values of the momentum it captures the profile quite well. On the other hand, 
DMFT improves the momentum dependence below $p_z < 2$~a.u., however it 
overestimates for values of the momentum in the range of 2~a.u. to 4~a.u.
The largest difference between the ``+U'' and ``+DMFT'' corrections are
seen along the [110] direction. This direction correspond to the shortest bond in the 
fcc structure. Here DMFT captures the peaks at around $0.5$~a.u. and the main 
peak at $2$~a.u., and continues very closely to the experimental data in 
the complete range of the computed momenta. LSDA+U captures only the main 
maximum and as a matter of fact produce worse results than LSDA. Along the 
$[111]$ direction DMFT get closer to the maximum at $p_z = 0$~a.u. than LSDA+U. 
For $p_z > 0$~a.u., the dip at $0.5$~a.u. is captured better within DMFT, 
while for higher momenta both the LSDA+U and LSDA+DMFT approaches follow 
essentially the same behavior. 

Although both static and dynamic corrections to the MCP spectra are rather 
similar, we observe a clear tendency of LSDA+U to overestimate the 
experimental data 
while LSDA+DMFT correct some discrepancies. The following subsection presents 
the results for the difference in total Compton profiles with respect to the
LSDA results, where distinctions because of static and dynamic corrections
became more apparent.

\subsection{Directional differences of Compton profiles}
\label{sec:cp_diff}
\begin{figure}[h]
   \includegraphics[width=0.99\linewidth,clip=true]{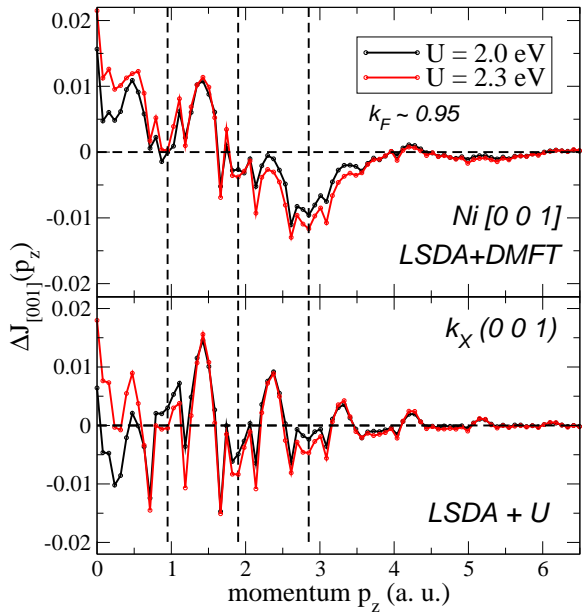}
\caption {\label{Fig:figure2}(color on-line) Computed difference Compton profiles of Ni
along the $[001]$ direction for different values of $U$. The difference spectra of LSDA+DMFT
with respect to LSDA, $U = 2.0$~eV (black-dashed), $U = 2.3$~eV (red-solid).}
\end{figure}

Fig. \ref{Fig:figure2} shows the total Compton profiles differences 
computed according to Eq. (\ref{mcs_dmft-lsda}) along the $[001]$ direction.  
The upper/lower part represents the DMFT/LSDA+U spectra after subtraction 
of LSDA results. 
%
One can easily recognize common features in comparing the LSDA+U with 
LSDA+DMFT spectra. The Brillouin zone boundary  along the [001] direction is 
represented by the $X(1/2, 0, 1/2)$-high symmetry point. 
The zone boundary is marked with the first dashed line and 
corresponds to the value of $k_F$ = 0.95 $a.u.$. 
The second dashed line is situated at $2k_F$ and is plotted to facilitate the
comparison between the spectra.
As one can see the LSDA+U spectra is sharper, 
since the DMFT self-energy contributes in smoothing out the spectra,
however the peaks remain in the same positions. 
No additional broadening of the spectra has been applied.
As a consequence of dynamic 
correlations within the first Brillouin zone, the $\Delta J_{[001]}$ has
positive weight, on contrary to the LSDA+U results.
We observe the Umklapp features
identified in the MCP spectra of Ni [001] in several studies 
(at $\sim$ 1.2, 1.7, 2.7 and 3.7 a.u.)
\cite{DDG+98,K04,KKK+03,BZK00} appearing in the $\Delta J_{[001]}$  as sharp
deeps. For the region with $1.9 < p_z < 3.8$~a.u. (third and fourth Brillouin 
zone), electronic correlations produces negative difference weights. Similar
 observation can be made for the different values of $U$. 

\begin{figure}[h]
   \includegraphics[width=0.99\linewidth]{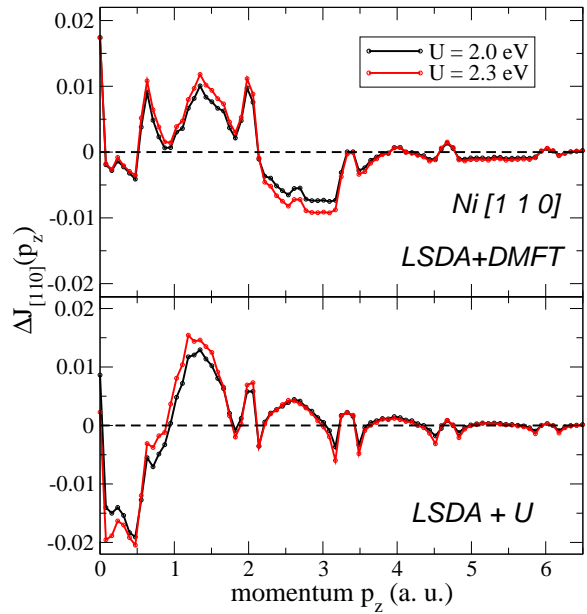}
\caption {\label{Fig:figure3}(color on-line) Computed difference Compton 
profiles of Ni along the [110] direction for different values of $U$. The 
difference spectra of LSDA+DMFT with respect to LSDA, $U=2.0~eV$ 
(black-dashed), $U=2.3~eV$ (red-solid).}
\end{figure}

Along the $[110]$ direction the Brillouin zone is intercepted at the $K$-point
with the coordinate 
$3\pi/2a(1, 1, 0)$ in the Cartesian representation. 
Similarly to the $[001]$ direction, the position of the main peaks of the 
spectra are the same in the LSDA+U/DMFT calculations. The Umklapp features 
identified in the MCP spectra by several studies at $\sim 2.0$, $3.3$ and
$4.6$~a.u. \cite{DDG+98,K04,KKK+03,BZK00} correspond  
to sharper peaks in the $\Delta J_{[110]}$ spectra. Again in LSDA+DMFT the 
spectra have an overall positive weight, due to correlation induced
life time effects determined by the imaginary part of the self-energy.
Also, Umklapp features are more visible in LSDA+U as no additional broadening 
is present.

\begin{figure}[h]
   \includegraphics[width=0.99\linewidth]{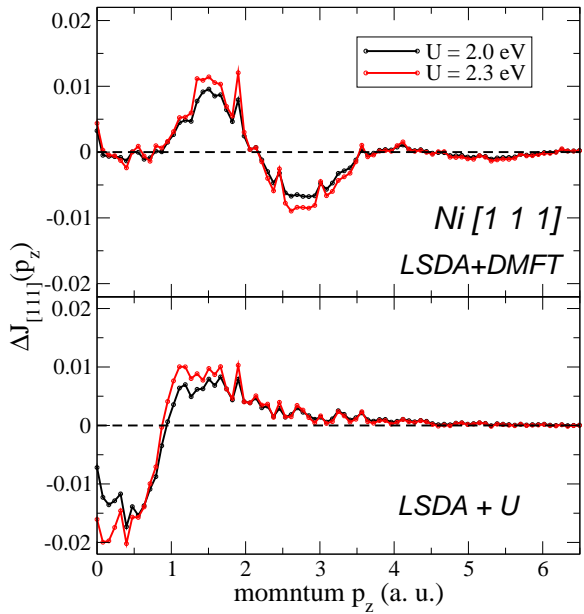}
\caption {\label{Fig:figure4}(color on-line) Computed difference Compton 
profiles of Ni along the $[111]$ direction for different values of $U$. The 
difference spectra of LSDA+DMFT with respect to LSDA, $U = 2.0$~eV 
(black-dashed), $U = 2.3$~eV (red-solid).}
\end{figure}

The same analysis can be performed upon the spectra in the $[111]$
direction. For  
momenta $p_z < k_F$ within the first Brillouin zone, $\Delta J_{[111]}$
computed with DMFT/+U have a negative contribution, whilst for the second
and third zone the  
LSDA+DMFT difference spectra have weights with alternating sign. The  
$\Delta J_{[111]}$  spectra are essentially negative only within the 
first zone and from $p_z > 1 a.u.$ the LSDA+U spectra is positive.

The general tendency in both LSDA+U and LSDA+DMFT is that for larger $U$, 
maxima and minima are slightly stretched out, while the modulations remain 
the same. This is expected as the modulation of the Compton profile is connected 
to the topology of the Fermi surface. As was previously demonstrated, cuts 
of the momentum density remain unchanged with inclusion of correlation
effects~\cite{M65}. The overall change in the shape of the Compton spectra 
comparing DMFT versus LSDA/LSDA+U reflects the presence of the imaginary 
part of the dynamic local self-energy. 

The comparison between the Compton spectra taken for different U values, 
along the same direction {\bf K} allows to discuss the strength of 
{\it local}-correlation effects.  In the same time comparing spectra 
obtained for a fixed $U$ value along different {\bf K}, {\bf K$^\prime$} 
directions
may reveal possible {\it non-local}-correlations effects. Although DMFT 
supplement the DFT-LDA part by a complex local and dynamic self-energy 
$\Sigma(z)$, the charge self-consistency of LDA+DMFT achieves indirect 
non-local effects. In the next subsection we analyze the differences 
between pairs of directional profiles also known as 
Compton profile anisotropies.

\subsection{Compton profiles anisotropy and non-locality}
\begin{figure}[h]
   \includegraphics[width=0.95\linewidth]{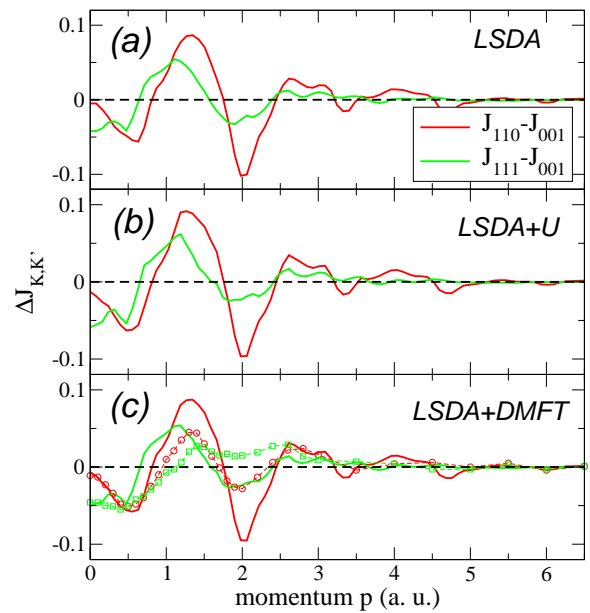}
\caption {\label{Fig:figure5}(color on-line) Theoretical Compton profile 
anisotropies of Ni. Upper panel (a) the anisotropy computed within LSDA. 
Lower pannels (b) and (c) show the LSDA+U and LSDA+DMFT results for 
$U = 2.3$~eV, $J = 0.9$~eV and $T = 400$~K. Comparison with the experimental 
anisotropy is presented in pannel (c).}
\end{figure}

Comparison between theoretical and experimental amplitudes of the Compton
profile anisotropies for Ni have already been 
performed~\cite{ei.re.74,wa.ca.75,AP90}.
The computed anisotropy profiles $J_{[110]} - J_{[001]}$ and  
$J_{[111]} - J_{[001]}$ are shown in Fig.~\ref{Fig:figure5} that, in 
contrast with earlier work, reveal the role of correlations effects.
In the upper panel the LSDA results are presented, while in the middle 
and lower panel the spectra obtained using LSDA+U and DMFT are seen. 
A very similar behavior, independent of the various level of sophistication 
to include the Coulomb interaction is visible.  
%

\begin{figure}[h]
   \includegraphics[width=0.95\linewidth]{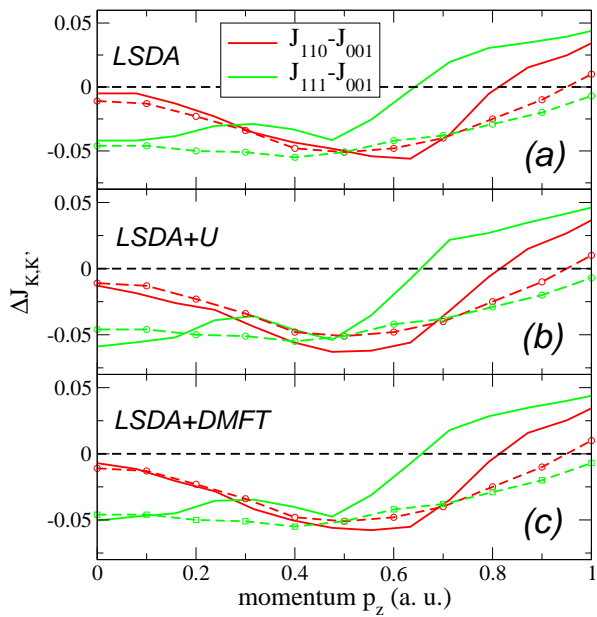}
\caption {\label{Fig:figure6}(color on-line) Theoretical Compton profile
anisotropies of Ni in comparison with the experimental measurements. 
LSDA/LSDA+U results, panel (a)/(b) over/under-estimate the anisotropy 
data. DMFT results are in a good agreement with the experimental 
spectra in the range $0 < p_z < 0.2$~a.u.
LSDA+U and LSDA+DMFT parameters are $U = 2.3$~eV, $J = 0.9$~eV and $T = 400$~K.}
\end{figure}

In the lower panel (c) we compare the anisotropy 
spectra with the corresponding experimental data of 
Anastassopoulos et. al. \cite{AP90}. 
There is a rather satisfying agreement between the theory and experiment, 
in particular for the difference $J_{[110]} - J_{[001]}$ 
(Fig.\ \ref{Fig:figure5} red lines) where the theoretical 
calculation follow most of the maxima and minima seen in the experiment.
Here again no broadening has been used for the  computed data. 
On the other hand for $J_{[111]} - J_{[001]}$ (Fig.~\ref{Fig:figure5} 
green lines) 
differences may be seen not only in the amplitude of the oscillation but 
also in the position of the minima/maxima. In Fig.\ \ref{Fig:figure6} we 
show the comparison on a reduced momentum regime $0 < p_z < 1$~a.u. In the 
upper pannel of Fig.\ \ref{Fig:figure6}(a), the LSDA results are seen to 
overestimate in the range $0 < p_z < 0.2$~a.u. the experimental spectra
plotted with dased lines. In the same momentum range the LSDA+U results 
underestimate the experimental data as seen in Fig.\ \ref{Fig:figure6}(b).
Figure\ \ref{Fig:figure6}(c) shows the LSDA+DMFT results and one can see 
that the dynamic correlations capture at best the behaviour of anisotropy
of the Compton profile in the region around the zero momentum $p_z < 0.2$~a.u.

Previous analysis attributed the discrepancies to the non-local correlation 
effects~\cite{RS87,AP90}, although no quantitative evidence has been presented. 
One possible alternative explanation for the results seen in 
Fig.\ \ref{Fig:figure6}
is that the density functional exchange-correlation potentials misplace the 
position of $d$-bands(orbitals). This agrees with the observation that LSDA 
overestimate the exchange splitting. Including static corrections using LSDA+U
the exchange splitting is enhanced and therefore this does not correct upon the 
position of the $d$-bands, and equally does not improve on the anisotropy 
spectra. On the contrary, LSDA+DMFT is known to improve on the exchange 
splitting as a consequence
of a Fermi-liquid type of self-energy and we equally see in the 
pannel of Fig.\ \ref{Fig:figure6}(c) although in a narrow momentum region 
$0 < p_z < 0.2$~a.u., an excellent agreement with the experimental 
anisotropies. 
As anisotropies in the Compton profile measures differences 
$J_{\bf K} - J_{\bf K'}$ they indicate non-local effects, therefore 
Fig.\ \ref{Fig:figure6}(c) shows that 
local dynamic correlations may capture non-locality in a very narrow  
region around zero momentum as a consequence of full change self-consitency of
LSDA+DMFT. In addition our results show that a description of the 
{\it non-local}-correlation effects is needed for a fruther improvement on the 
amplitudes of Compton profiles anisotropies for larger momenta. 

\section{Kinetic energy corrections from relativistic and electronic
correlations}\label{sec:ke}
The relativistic generalization of the Schr\"odinger theory of quantum 
mechanics to describe particles with spin 1/2 was achieved using the  
Dirac equation (see for example~\cite{T3}). The construction of Dirac 
equation uses symmetry arguments and energy considerations, and it 
starts from the general Hamiltonian:
\begin{equation}\label{ham_D}
H=  \frac{c}{i} {\bf \alpha} \cdot {\bf \nabla} + 
\frac{1}{2} \left( \beta -I   \right) +  V({\bf r}),
\end{equation}
where ${\bf \alpha}$ and $\beta$ are standard Dirac matrices and 
$V({\bf r})$ represents a one particle effective potential. In the 
spirit of the non-relativistic DFT the effective
potential consists of the Hartree term, an exchange correlation and 
a spin dependent part~\cite{eb.fr.97,E00}: 
$V({\bf r})= V_H({\bf r})+V_{xc}({\bf r})+ \beta \sigma_z B({\bf r})$.
Rigorous four-component relativistic many-electron calculations are hardly 
tractable in the spirit of four-component Dirac relativistic quantum 
mechanics~\cite{D28}. 
It is important to 
note that many applications consider relativistic effects at one-component
level, which is frequently called the scalar relativistic approach~\cite{FW50} 
which usually assume that corrections of higher order than $1/c^2$ can be 
neglected for chemical accuracy. Furthermore it is assumed that the effect of 
the spin-orbit coupling on the form of the orbitals may be neglected, allowing 
a partition of the Hamiltonian into a spin-independent and a spin-dependent 
part. The latter part is then only used in the final stage of the calculation 
to couple to the correlated many-electron problem. The analysis of components 
of the spin-orbit coupling was performed decoupling the longitudinal and 
transversal contributions ~\cite{eb.fr.97}, allowing to identify the source 
of the most important spin-orbit induced phenomena in solids. 

There are a few methods available to quantitatively assess the interplay
between correlation and relativistic effects. Within the framework of
DFT the recently developed LSDA+DMFT scheme demonstrates a 
clear potential in this direction. LSDA+DMFT has been systematically applied 
to $d$-, $f$-electron systems with various DMFT solvers~\cite{KS06}. 
From a pragmatic point of view perturbative solvers of DMFT written 
in adapted basis sets to include spin-orbit effects \cite{PKL05} are 
efficient tools for realistic multi atom/orbital calculations. 
This means that we in fact capture the interplay between relativistic 
and correlation effects at a more economical level of the theory: from 
the correlations point of view a perturbative solver is considered, while 
for the relativistic part the four-component was replaced by a two 
component formulation.
For a single-particle in an effective potential $V_{eff}$, the most common 
transformation of the Dirac Hamiltonian (Eq.~(\ref{ham_D})) into the two 
component formulation: $H_{2comp} = U H_{4comp}U^{\dagger}$ is expressed 
as a unitary transformation~\cite{FW50}, followed by a Taylor expansion in 
the fine structure constant $\propto 1/c^2$ and produce the following terms:
\begin{eqnarray}\label{BP}
H^{BP} &=& (mc^2 +V_{eff} + \frac{p^2}{2m}) - \frac{p^4}{8m^3c^2} \\
&-& \frac{1}{8m^2c^2}(p^2 V_{eff}) + \frac{\hbar}{4m^2c^2}{\bf \sigma}(\nabla V_{eff} \times \vec{p}) + ... \nonumber
\end{eqnarray}
The first terms in parenthesis in Eq. (\ref{BP}) represent the usual 
non-relativistic Hamiltonian, then the second one is the so called 
mass-velocity term, the third is called the Darwin term and the fourth 
operator describes the spin-orbit coupling (interaction). 
It can be analytically proved that the scalar mass-velocity and Darwin 
terms are unbounded from below.  The resulting Breit-Pauli 
Hamiltonian ($H^{BP}$) also known as the first order relativistic 
Hamiltonian contain terms that are highly-singular and variationally
instable. Therefore this operator is suitable to be used in 
the low order perturbation theory.

\subsection{Moments of the differences of the Compton profiles}
The measured Compton profile, or the momentum distribution enable in 
principle to obtain averages $\langle p^n \rangle$ directly from the 
experiment. On the computational side, the momentum space formulation 
allows to obtain the Compton profiles within the LSDA~\cite{C85} and 
LSDA+DMFT~\cite{BMC+12,CBE+14} for many systems. 
Further on additional information can be gained by taking moments of 
the difference between the correlated and non-correlated Compton profiles
along different {\bf K}-directions: $p_z || {\bf K}$
\begin{equation}
\langle p^n \rangle_{\bf K} = \int_0^{\infty} p_z^n \left[ J^{+U/DMFT}_{\bf K}(p_z) - 
J^{LSDA}_{\bf K}(p_z) \right] dp_z;   \nonumber 
\end{equation}
Recently we have computed the second moments $\langle p^2 \rangle$ 
~\cite{CBE+14} along specific directions in Fe and Ni and discussed the effect 
of electronic correlations upon the kinetic energy per bonds. Aside from the 
kinetic energy $\langle p^2 \rangle$, it is possible to obtain also 
relativistic energy corrections in the form of $\langle p^4 \rangle$. 
In the present work we are interested to estimate the relativistic corrections 
to the kinetic energy that arises from the variation of electron mass with 
velocity, and how it may vary as a function of the local Coulomb interaction. 

The second and the forth moments of the difference in the total Compton 
profiles, along the bond directions would provide some specific terms from 
the expansion (\ref{BP}). Namely we are going to evaluate the so called free 
particle relativistic kinetic energy ($H_0$) in terms of the second moments 
and its relativistic correction as the forth moment, along the bond 
directions of Ni. 
In order to compare the magnitude of second and fourth order moments one has 
to introduce a dimensionless quantity $p_r=p/mc$. In this reduced variable 
$H_0 = mc^2 \left( (1/2) p_r^2  - (1/8) p_r^4 +... \right) $
and the relevant expectation values has the expression:
\begin{eqnarray}
\langle H_0 \rangle_{\bf K} &=& \int H_0 \Delta J^{+U/DMFT}_{\bf K}(p_z) dp_z;   (p_z || {\bf K}) \nonumber \\
& \approx & mc^2 \left[ \frac{1}{2} \langle p_r^2 \rangle_{\bf K} 
- \frac{1}{8} \langle p_r^4 \rangle_{\bf K} \right] \label{ho_k}
\end{eqnarray}

\begin{table}[h]
\caption{\label{tab1}
Expectation values $\langle p_r^n\rangle$ along different direction  computed within LSDA+U}
\begin{ruledtabular}
\begin{tabular}{ccccc}
               &     & \multicolumn{3}{c}{LSDA+U}      \\ \hline
     $n$        & $U$   & [001]     & [110]     & [111]   \\ 
               & eV  & 10$^{-6}$ & 10$^{-6}$ & 10$^{-6}$ \\ \hline
      2        & 2.0 &  1.06     &  2.65     & 6.37   \\
               & 2.3 & -2.12     &  0.53     & 5.84   \\
      4        & 2.0 & -4.20 10$^{-3}$  & -1.18 10$^{-3}$  & 4.26 10$^{-3}$ \\
               & 2.3 & -6.29 10$^{-3}$  & -4.46 10$^{-3}$  & 2.43 10$^{-3}$ \\ 
\end{tabular}
\end{ruledtabular}
\end{table}
In the momentum space representation such integrals can be directly 
computed. We performed calculations for the second and fourth moments for 
different values of $U$ and different level of electronic correlations.
The LSDA+U results are given in the Table \ref{tab1}. One can clearly see 
that the second and fourth order moments differ significantly along different 
directions. For $U = 2$~eV all second moments are positive for all directions, 
while for larger $U = 2.3$~eV along [001] the second moment is negative. In 
the reduced representation $p/mc$ the magnitude of these moments is of order 
$10^{-6}$. The forth order moments are 3 order of magnitude smaller than the 
second order moments, and are negative along the [001] and [110] 
directions. Along the [111] direction the forth order moment remain
positive for all $U$ values. 
Note that moments decrease 
in magnitude as the distance is increasing: the largest moments
are obtained for the nearest neighbors distances, which are the 
shortest bonds. For Ni, this corresponds to the [110] direction.
Directional averaging over all second order moments provides the 
kinetic energy, while a similar average over the 4-th order moments 
provides the relativistic corrections to the kinetic energy. 

The moments computed in LSDA+DMFT are given in  Table \ref{tab2}.
On contrary to the LSDA+U, using DMFT produces a Compton profile having 
negative second moments along all directions and for all studied values 
of $U = 2$~eV and $2.3$~eV.

\begin{table}[h]
\caption{\label{tab2}
Expectation values $\langle p_r^n \rangle$ along different direction 
computed within LSDA+DMFT}
\begin{ruledtabular}
\begin{tabular}{ccccc}
               &     & \multicolumn{3}{c}{LSDA+DMFT}   \\ \hline
     $n$        & $U$   & [001]     & [110]     & [111]   \\
               & eV  & 10$^{-6}$ & 10$^{-6}$ & 10$^{-6}$ \\ \hline
      2        & 2.0 & -7.43     & -6.90     & -4.78  \\
               & 2.3 &-11.15     & -8.50     & -6.37  \\
      4        & 2.0 &  8.75 10$^{-3}$  & -5.92 10$^{-3}$  & -3.69 10$^{-3}$ \\
               & 2.3 & -1.12 10$^{-3}$  & -7.62 10$^{-3}$  & -6.38 10$^{-3}$ \\
\end{tabular}
\end{ruledtabular}
\end{table}

\begin{figure}[h]
   \includegraphics[width=0.99\linewidth, clip=true]{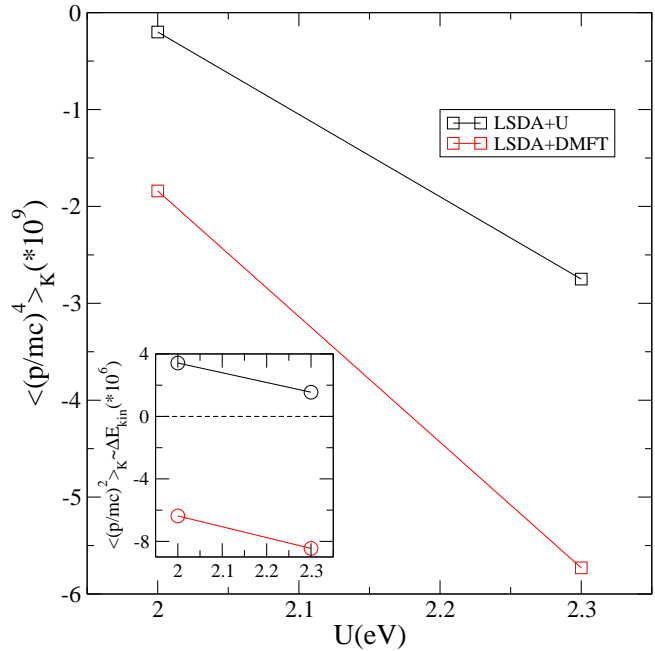}
\caption {\label{Fig:figure7}(color on-line) Averaged forth (second, in inset) 
order moments of the difference Compton profiles computed within DMFT 
(red solid) and LSDA+U (black dashed), along different directions and 
for the Coulomb parameters $U = 2$~eV$/2.3$~eV, $J = 0.9$~eV and $T = 400$~K.}
\end{figure}

Note the qualitative difference between the LSDA+U and LSDA+DMFT second 
moments: while the former mean field (LSDA+U) approach produce second moments 
with different signs depending on the directions, within the later dynamic 
(LSDA+DMFT) approach correction is always negative. Similarly to our previous 
results~\cite{CBE+14} we see that the positive difference at low momentum 
region $0 < p_z < 2$~a.u. is completely overruled by the negative weights at 
higher momenta, which leads to the overall negative values for the correction 
obtained in DMFT.
In LSDA+U a negative second moment is obtained along [001] and positive for 
the other two directions. The positive second moment is obtained as the 
Compton profile computed in the mean field (LSDA+U) approach always is larger 
than the corresponding LSDA profile, for any value of the moment $p_z$.  
In order to discuss correction to the kinetic energies,
we computed the weighted sum of the nearest neighbors, i.e.\ six times the 
contribution along [001], 12 times the contribution along [110] and 8 times 
the contribution along [111] divided by the total number of neighbors (26).
Fig. \ref{Fig:figure7} summarizes the computed results. The
inset shows the directional average of the second moment which 
is positive in LSDA+U and negative in DMFT, whilst the corrections 
to the kinetic energy is negative in both +U/DMFT calculations.

\section{Discussions and Conclusion}
\label{sec:conc}
The influence of electronic correlations on the Compton profiles of Ni
has been discussed within the framework of DFT 
comparing the results of mean field LSDA+U and beyond mean-field LSDA+DMFT. 
According to our results, the mean field decoupling of the interaction (+U) 
overestimates slightly the MCP spectra, while dynamic correlations 
improve the agreement with experiment. To reveal differences between 
the LSDA+U and LSDA+DMFT approaches we studied the directional differences, 
i.e. differences of Compton profiles with respect to the LSDA spectra.
Overall the difference spectra follow a similar momentum dependence with
visible deviations in the low momentum region. A qualitative difference 
is evidenced in this region: within the mean field approach (+U) negative
differences are seen while in the dynamic case, the opposite result is 
obtained. In other words, the mean field LSDA+U Compton spectra have
a smaller weight in the low energy region than the corresponding 
LSDA+DMFT Compton spectra. According to our recent picture of momentum 
redistribution because of interaction~\cite{CBE+14} we conclude that the 
weight from low momentum distribution is shifted towards
the higher momentum region in the LSDA+U spectra. This is in 
agreement with the naive picture of the effects of LSDA+U on the spectral 
weight distribution shifting weights towards higher energies. In the
Compton scattering language, photons would scatter accordingly  
on moving electrons situated in higher energy bands, although this does 
not mean that the electrons are moving faster, explaining the fact 
that there are no dramatic changes in the Compton spectra (differences 
of order of $\pm 0.02$) shown in Figs.\ \ref{Fig:figure2}, \ref{Fig:figure3} 
and \ref{Fig:figure4}. On the contrary to the LSDA+U results, in the DMFT 
calculations the Fermi liquid type of self-energy determines the spectral 
weight transfer towards the low energy region, and accordingly the spectra 
of photons scattering on the renormalized electronic structure would be 
redistributed towards low momenta. Similar conclusions have been reached 
in our previous studies~\cite{BMC+12,CBE+14}. 

In the analysis of the Compton profile anisotropies we found that the 
LSDA+DMFT results describe well the momentum region of 
$0 < p_z < 0.2$~a.u. which is a consequence of the presence of a local 
and dynamic self-energy that properly locates the position of Ni $d$-bands,
as seen in various calculations. 
The limited momentum range is due to the inherent DMFT approximation 
that the self-energy neglect spatial fluctuations. 

In order to assess the capability of electronic correlations to influence 
the kinetic energy along specific directions we have computed the second 
and the fourth order moments of the Compton spectra considering the reduced 
momentum $p_r = p/mc$. Although the fourth order moments are significantly 
smaller, $\langle p_r^4 \rangle \propto 10^{-3} \langle p_r^2 \rangle $,
an overall non-negligible contribution is obtained (see Eq.(\ref{ho_k})). 
Within LSDA+U the second moment has a positive sign, which is in agreement 
with the description of the momentum redistribution towards higher momenta
descried above, except for the [001] direction. The overall energy correction
is still positive in LSDA+U as seen in the inset of Fig.\ \ref{Fig:figure7}. 
Negative second moments are obtained along all directions in DMFT and produce
a negative kinetic energy contribution. The relativistic corrections to the
kinetic energy are both negative and we see that dynamic correlations 
(LSDA+DMFT) generate larger relativistic corrections to the one-particle 
kinetic energy in comparison to their mean field (LSDA+U) counter part. 

As an overall conclusion in the range of the studied values of $U$
qualitative and quantitative differences are seen in the Compton 
profiles depending weather the LSDA is supplemented with static or 
dynamic many-body effects. An important message is that relativistic 
effects and electronic correlations may have a non-trivial interplay and 
dynamic correlations determine larger relativistic corrections in the 
electronic structure of solids. Further investigations are necessary 
for a quantitative assessments of such effects. 

\section{acknowledgments}
Financial support of the
Deutsche Forschungsgemeinschaft through FOR 1346, the DAAD and the CNCS - UEFISCDI 
(project number PN-II-ID-PCE-2012-4-0470) is gratefully acknowledged.
JM acknowledge also the CENTEM project, reg. no. CZ.1.05/2.1.00/03.0088, co-funded 
by the ERDF as part of the Ministry of Education, Youth and Sports OP RDI programme.
    

\bibliography{paper}

\begin{thebibliography}{59}
\expandafter\ifx\csname natexlab\endcsname\relax\def\natexlab#1{#1}\fi
\expandafter\ifx\csname bibnamefont\endcsname\relax
  \def\bibnamefont#1{#1}\fi
\expandafter\ifx\csname bibfnamefont\endcsname\relax
  \def\bibfnamefont#1{#1}\fi
\expandafter\ifx\csname citenamefont\endcsname\relax
  \def\citenamefont#1{#1}\fi
\expandafter\ifx\csname url\endcsname\relax
  \def\url#1{\texttt{#1}}\fi
\expandafter\ifx\csname urlprefix\endcsname\relax\def\urlprefix{URL }\fi
\providecommand{\bibinfo}[2]{#2}
\providecommand{\eprint}[2][]{\url{#2}}

\bibitem[{\citenamefont{Aubert and Michelutti}(1977)}]{AM77}
\bibinfo{author}{\bibfnamefont{G.}~\bibnamefont{Aubert}} \bibnamefont{and}
  \bibinfo{author}{\bibfnamefont{B.}~\bibnamefont{Michelutti}},
  \bibinfo{journal}{Physica B} \textbf{\bibinfo{volume}{86-88}},
  \bibinfo{pages}{295} (\bibinfo{year}{1977}).

\bibitem[{\citenamefont{Yang et~al.}(2001)\citenamefont{Yang, Savrasov, and
  Kotliar}}]{YS01}
\bibinfo{author}{\bibfnamefont{I.}~\bibnamefont{Yang}},
  \bibinfo{author}{\bibfnamefont{S.~Y.} \bibnamefont{Savrasov}},
  \bibnamefont{and} \bibinfo{author}{\bibfnamefont{G.}~\bibnamefont{Kotliar}},
  \bibinfo{journal}{Phys. Rev. Letters} \textbf{\bibinfo{volume}{87}},
  \bibinfo{pages}{216405} (\bibinfo{year}{2001}).

\bibitem[{\citenamefont{Gersdorf}(1978)}]{G78}
\bibinfo{author}{\bibfnamefont{R.}~\bibnamefont{Gersdorf}},
  \bibinfo{journal}{Phys. Rev. Letters} \textbf{\bibinfo{volume}{40}},
  \bibinfo{pages}{344} (\bibinfo{year}{1978}).

\bibitem[{\citenamefont{B\"unemann et~al.}(2008)\citenamefont{B\"unemann,
  Gebhard, Ohm, Weiser, and Weber}}]{BG08}
\bibinfo{author}{\bibfnamefont{J.}~\bibnamefont{B\"unemann}},
  \bibinfo{author}{\bibfnamefont{F.}~\bibnamefont{Gebhard}},
  \bibinfo{author}{\bibfnamefont{T.}~\bibnamefont{Ohm}},
  \bibinfo{author}{\bibfnamefont{S.}~\bibnamefont{Weiser}}, \bibnamefont{and}
  \bibinfo{author}{\bibfnamefont{W.}~\bibnamefont{Weber}},
  \bibinfo{journal}{Phys. Rev. Letters} \textbf{\bibinfo{volume}{101}},
  \bibinfo{pages}{236404} (\bibinfo{year}{2008}).

\bibitem[{\citenamefont{Leung et~al.}(1991)\citenamefont{Leung, Chan, and
  Harmon}}]{LC91}
\bibinfo{author}{\bibfnamefont{T.~C.} \bibnamefont{Leung}},
  \bibinfo{author}{\bibfnamefont{C.~T.} \bibnamefont{Chan}}, \bibnamefont{and}
  \bibinfo{author}{\bibfnamefont{B.~N.} \bibnamefont{Harmon}},
  \bibinfo{journal}{Phys. Rev. B} \textbf{\bibinfo{volume}{44}},
  \bibinfo{pages}{2923} (\bibinfo{year}{1991}).

\bibitem[{\citenamefont{Eastman et~al.}(1978)\citenamefont{Eastman, Himpsel,
  and Knapp}}]{EH78}
\bibinfo{author}{\bibfnamefont{D.~E.} \bibnamefont{Eastman}},
  \bibinfo{author}{\bibfnamefont{F.~J.} \bibnamefont{Himpsel}},
  \bibnamefont{and} \bibinfo{author}{\bibfnamefont{J.~A.} \bibnamefont{Knapp}},
  \bibinfo{journal}{Phys. Rev. Lett.} \textbf{\bibinfo{volume}{40}},
  \bibinfo{pages}{1514} (\bibinfo{year}{1978}).

\bibitem[{\citenamefont{Dietz et~al.}(1978)\citenamefont{Dietz, Gerhardt, and
  Maetz}}]{DG78}
\bibinfo{author}{\bibfnamefont{E.}~\bibnamefont{Dietz}},
  \bibinfo{author}{\bibfnamefont{U.}~\bibnamefont{Gerhardt}}, \bibnamefont{and}
  \bibinfo{author}{\bibfnamefont{C.~J.} \bibnamefont{Maetz}},
  \bibinfo{journal}{Phys. Rev. Lett.} \textbf{\bibinfo{volume}{40}},
  \bibinfo{pages}{892} (\bibinfo{year}{1978}).

\bibitem[{\citenamefont{Himpsel et~al.}(1979)\citenamefont{Himpsel, Knapp, and
  Eastman}}]{HK79}
\bibinfo{author}{\bibfnamefont{F.~J.} \bibnamefont{Himpsel}},
  \bibinfo{author}{\bibfnamefont{J.~A.} \bibnamefont{Knapp}}, \bibnamefont{and}
  \bibinfo{author}{\bibfnamefont{D.~E.} \bibnamefont{Eastman}},
  \bibinfo{journal}{Phys. Rev. B} \textbf{\bibinfo{volume}{19}},
  \bibinfo{pages}{2919} (\bibinfo{year}{1979}).

\bibitem[{\citenamefont{Eastman et~al.}(1980)\citenamefont{Eastman, Himpsel,
  and Knapp}}]{EH80}
\bibinfo{author}{\bibfnamefont{D.~E.} \bibnamefont{Eastman}},
  \bibinfo{author}{\bibfnamefont{F.~J.} \bibnamefont{Himpsel}},
  \bibnamefont{and} \bibinfo{author}{\bibfnamefont{J.~A.} \bibnamefont{Knapp}},
  \bibinfo{journal}{Phys. Rev. Lett.} \textbf{\bibinfo{volume}{44}},
  \bibinfo{pages}{95} (\bibinfo{year}{1980}).

\bibitem[{\citenamefont{Eberhardt and Plummer}(1980)}]{EP80}
\bibinfo{author}{\bibfnamefont{W.}~\bibnamefont{Eberhardt}} \bibnamefont{and}
  \bibinfo{author}{\bibfnamefont{E.~W.} \bibnamefont{Plummer}},
  \bibinfo{journal}{Phys. Rev. B} \textbf{\bibinfo{volume}{21}},
  \bibinfo{pages}{3245} (\bibinfo{year}{1980}).

\bibitem[{\citenamefont{Guillot et~al.}(1977)\citenamefont{Guillot, Bally,
  Paign{\'e}, Lecante, Jain, Thiry, Pinchaux, Petroff, and Falicov}}]{GBP+77}
\bibinfo{author}{\bibfnamefont{C.}~\bibnamefont{Guillot}},
  \bibinfo{author}{\bibfnamefont{Y.}~\bibnamefont{Bally}},
  \bibinfo{author}{\bibfnamefont{J.}~\bibnamefont{Paign{\'e}}},
  \bibinfo{author}{\bibfnamefont{J.}~\bibnamefont{Lecante}},
  \bibinfo{author}{\bibfnamefont{K.~P.} \bibnamefont{Jain}},
  \bibinfo{author}{\bibfnamefont{P.}~\bibnamefont{Thiry}},
  \bibinfo{author}{\bibfnamefont{R.}~\bibnamefont{Pinchaux}},
  \bibinfo{author}{\bibfnamefont{Y.}~\bibnamefont{Petroff}}, \bibnamefont{and}
  \bibinfo{author}{\bibfnamefont{L.~M.} \bibnamefont{Falicov}},
  \bibinfo{journal}{Phys. Rev. Letters} \textbf{\bibinfo{volume}{39}},
  \bibinfo{pages}{1632} (\bibinfo{year}{1977}).

\bibitem[{\citenamefont{Metzner and Vollhardt}(1989)}]{MV89}
\bibinfo{author}{\bibfnamefont{W.}~\bibnamefont{Metzner}} \bibnamefont{and}
  \bibinfo{author}{\bibfnamefont{D.}~\bibnamefont{Vollhardt}},
  \bibinfo{journal}{Phys. Rev. Letters} \textbf{\bibinfo{volume}{62}},
  \bibinfo{pages}{324} (\bibinfo{year}{1989}).

\bibitem[{\citenamefont{Georges et~al.}(1996)\citenamefont{Georges, Kotliar,
  Krauth, and Rozenberg}}]{GK96}
\bibinfo{author}{\bibfnamefont{A.}~\bibnamefont{Georges}},
  \bibinfo{author}{\bibfnamefont{G.}~\bibnamefont{Kotliar}},
  \bibinfo{author}{\bibfnamefont{W.}~\bibnamefont{Krauth}}, \bibnamefont{and}
  \bibinfo{author}{\bibfnamefont{M.~J.} \bibnamefont{Rozenberg}},
  \bibinfo{journal}{Rev. Mod. Phys.} \textbf{\bibinfo{volume}{68}},
  \bibinfo{pages}{13} (\bibinfo{year}{1996}).

\bibitem[{\citenamefont{Kotliar and Vollhardt}(2004)}]{KV04}
\bibinfo{author}{\bibfnamefont{G.}~\bibnamefont{Kotliar}} \bibnamefont{and}
  \bibinfo{author}{\bibfnamefont{D.}~\bibnamefont{Vollhardt}},
  \bibinfo{journal}{Phys. Today} \textbf{\bibinfo{volume}{57}},
  \bibinfo{pages}{53} (\bibinfo{year}{2004}).

\bibitem[{\citenamefont{Lichtenstein et~al.}(2001)\citenamefont{Lichtenstein,
  Katsnelson, and Kotliar}}]{LK01}
\bibinfo{author}{\bibfnamefont{A.~I.} \bibnamefont{Lichtenstein}},
  \bibinfo{author}{\bibfnamefont{M.~I.} \bibnamefont{Katsnelson}},
  \bibnamefont{and} \bibinfo{author}{\bibfnamefont{G.}~\bibnamefont{Kotliar}},
  \bibinfo{journal}{Phys. Rev. Letters} \textbf{\bibinfo{volume}{87}},
  \bibinfo{pages}{067205} (\bibinfo{year}{2001}).

\bibitem[{\citenamefont{Chioncel et~al.}(2003)\citenamefont{Chioncel, Vitos,
  Abrikosov, Koll\'ar, Katsnelson, and Lichtenstein}}]{CVA+03}
\bibinfo{author}{\bibfnamefont{L.}~\bibnamefont{Chioncel}},
  \bibinfo{author}{\bibfnamefont{L.}~\bibnamefont{Vitos}},
  \bibinfo{author}{\bibfnamefont{I.~A.} \bibnamefont{Abrikosov}},
  \bibinfo{author}{\bibfnamefont{J.}~\bibnamefont{Koll\'ar}},
  \bibinfo{author}{\bibfnamefont{M.~I.} \bibnamefont{Katsnelson}},
  \bibnamefont{and} \bibinfo{author}{\bibfnamefont{A.~I.}
  \bibnamefont{Lichtenstein}}, \bibinfo{journal}{Phys. Rev. B}
  \textbf{\bibinfo{volume}{67}}, \bibinfo{pages}{235106}
  (\bibinfo{year}{2003}).

\bibitem[{\citenamefont{Min\'ar et~al.}(2005)\citenamefont{Min\'ar, Chioncel,
  Perlov, Ebert, Katsnelson, and Lichtenstein}}]{MCP+05}
\bibinfo{author}{\bibfnamefont{J.}~\bibnamefont{Min\'ar}},
  \bibinfo{author}{\bibfnamefont{L.}~\bibnamefont{Chioncel}},
  \bibinfo{author}{\bibfnamefont{A.}~\bibnamefont{Perlov}},
  \bibinfo{author}{\bibfnamefont{H.}~\bibnamefont{Ebert}},
  \bibinfo{author}{\bibfnamefont{M.~I.} \bibnamefont{Katsnelson}},
  \bibnamefont{and} \bibinfo{author}{\bibfnamefont{A.~I.}
  \bibnamefont{Lichtenstein}}, \bibinfo{journal}{Phys. Rev. B}
  \textbf{\bibinfo{volume}{72}}, \bibinfo{pages}{045125}
  (\bibinfo{year}{2005}).

\bibitem[{\citenamefont{Braun et~al.}(2006)\citenamefont{Braun, Min\'ar, Ebert,
  Katsnelson, and Lichtenstein}}]{BME+06}
\bibinfo{author}{\bibfnamefont{J.}~\bibnamefont{Braun}},
  \bibinfo{author}{\bibfnamefont{J.}~\bibnamefont{Min\'ar}},
  \bibinfo{author}{\bibfnamefont{H.}~\bibnamefont{Ebert}},
  \bibinfo{author}{\bibfnamefont{M.~I.} \bibnamefont{Katsnelson}},
  \bibnamefont{and} \bibinfo{author}{\bibfnamefont{A.~I.}
  \bibnamefont{Lichtenstein}}, \bibinfo{journal}{Phys. Rev. Lett.}
  \textbf{\bibinfo{volume}{97}}, \bibinfo{pages}{227601}
  (\bibinfo{year}{2006}).

\bibitem[{\citenamefont{Grechnev et~al.}(2007)\citenamefont{Grechnev, Marco,
  Katsnelson, Lichtenstein, Wills, and Eriksson}}]{GM07}
\bibinfo{author}{\bibfnamefont{A.}~\bibnamefont{Grechnev}},
  \bibinfo{author}{\bibfnamefont{I.~D.} \bibnamefont{Marco}},
  \bibinfo{author}{\bibfnamefont{M.~I.} \bibnamefont{Katsnelson}},
  \bibinfo{author}{\bibfnamefont{A.~I.} \bibnamefont{Lichtenstein}},
  \bibinfo{author}{\bibfnamefont{J.}~\bibnamefont{Wills}}, \bibnamefont{and}
  \bibinfo{author}{\bibfnamefont{O.}~\bibnamefont{Eriksson}},
  \bibinfo{journal}{Phys. Rev. B} \textbf{\bibinfo{volume}{76}},
  \bibinfo{pages}{035107} (\bibinfo{year}{2007}).

\bibitem[{\citenamefont{S\'anchez-Barriga
  et~al.}(2009)\citenamefont{S\'anchez-Barriga, Fink, Boni, Di~Marco, Braun,
  Min\'ar, Varykhalov, Rader, Bellini, Manghi et~al.}}]{SFB+09}
\bibinfo{author}{\bibfnamefont{J.}~\bibnamefont{S\'anchez-Barriga}},
  \bibinfo{author}{\bibfnamefont{J.}~\bibnamefont{Fink}},
  \bibinfo{author}{\bibfnamefont{V.}~\bibnamefont{Boni}},
  \bibinfo{author}{\bibfnamefont{I.}~\bibnamefont{Di~Marco}},
  \bibinfo{author}{\bibfnamefont{J.}~\bibnamefont{Braun}},
  \bibinfo{author}{\bibfnamefont{J.}~\bibnamefont{Min\'ar}},
  \bibinfo{author}{\bibfnamefont{A.}~\bibnamefont{Varykhalov}},
  \bibinfo{author}{\bibfnamefont{O.}~\bibnamefont{Rader}},
  \bibinfo{author}{\bibfnamefont{V.}~\bibnamefont{Bellini}},
  \bibinfo{author}{\bibfnamefont{F.}~\bibnamefont{Manghi}},
  \bibnamefont{et~al.}, \bibinfo{journal}{Phys. Rev. Lett.}
  \textbf{\bibinfo{volume}{103}}, \bibinfo{pages}{267203}
  (\bibinfo{year}{2009}).

\bibitem[{\citenamefont{Granas et~al.}(2012)\citenamefont{Granas, di~Marco,
  Thunstr\"om, Nordstr\"om, Eriksson, Bj\"orkman, and Wills}}]{GM12}
\bibinfo{author}{\bibfnamefont{O.}~\bibnamefont{Granas}},
  \bibinfo{author}{\bibfnamefont{I.}~\bibnamefont{di~Marco}},
  \bibinfo{author}{\bibfnamefont{P.}~\bibnamefont{Thunstr\"om}},
  \bibinfo{author}{\bibfnamefont{L.}~\bibnamefont{Nordstr\"om}},
  \bibinfo{author}{\bibfnamefont{O.}~\bibnamefont{Eriksson}},
  \bibinfo{author}{\bibfnamefont{T.}~\bibnamefont{Bj\"orkman}},
  \bibnamefont{and} \bibinfo{author}{\bibfnamefont{J.}~\bibnamefont{Wills}},
  \bibinfo{journal}{Comp. Mat. Sci.} \textbf{\bibinfo{volume}{55}},
  \bibinfo{pages}{295} (\bibinfo{year}{2012}).

\bibitem[{\citenamefont{S\'anchez-Barriga
  et~al.}(2012)\citenamefont{S\'anchez-Barriga, Braun, Min\'ar, Di~Marco,
  Varykhalov, Rader, Boni, Bellini, Manghi, Ebert et~al.}}]{SBM+12}
\bibinfo{author}{\bibfnamefont{J.}~\bibnamefont{S\'anchez-Barriga}},
  \bibinfo{author}{\bibfnamefont{J.}~\bibnamefont{Braun}},
  \bibinfo{author}{\bibfnamefont{J.}~\bibnamefont{Min\'ar}},
  \bibinfo{author}{\bibfnamefont{I.}~\bibnamefont{Di~Marco}},
  \bibinfo{author}{\bibfnamefont{A.}~\bibnamefont{Varykhalov}},
  \bibinfo{author}{\bibfnamefont{O.}~\bibnamefont{Rader}},
  \bibinfo{author}{\bibfnamefont{V.}~\bibnamefont{Boni}},
  \bibinfo{author}{\bibfnamefont{V.}~\bibnamefont{Bellini}},
  \bibinfo{author}{\bibfnamefont{F.}~\bibnamefont{Manghi}},
  \bibinfo{author}{\bibfnamefont{H.}~\bibnamefont{Ebert}},
  \bibnamefont{et~al.}, \bibinfo{journal}{Phys. Rev. B}
  \textbf{\bibinfo{volume}{85}}, \bibinfo{pages}{205109}
  (\bibinfo{year}{2012}).

\bibitem[{\citenamefont{Kubo and Asano}(1990)}]{KA90}
\bibinfo{author}{\bibfnamefont{Y.}~\bibnamefont{Kubo}} \bibnamefont{and}
  \bibinfo{author}{\bibfnamefont{S.}~\bibnamefont{Asano}},
  \bibinfo{journal}{Phys. Rev. B} \textbf{\bibinfo{volume}{42}},
  \bibinfo{pages}{4431} (\bibinfo{year}{1990}).

\bibitem[{\citenamefont{Dixon et~al.}(1998)\citenamefont{Dixon, Duffy,
  Gardelis, McCarthy, Cooper, Dugdale, Jarlborg, and Timms}}]{DDG+98}
\bibinfo{author}{\bibfnamefont{M.~A.~G.} \bibnamefont{Dixon}},
  \bibinfo{author}{\bibfnamefont{J.~A.} \bibnamefont{Duffy}},
  \bibinfo{author}{\bibfnamefont{S.}~\bibnamefont{Gardelis}},
  \bibinfo{author}{\bibfnamefont{J.~E.} \bibnamefont{McCarthy}},
  \bibinfo{author}{\bibfnamefont{M.~J.} \bibnamefont{Cooper}},
  \bibinfo{author}{\bibfnamefont{S.~B.} \bibnamefont{Dugdale}},
  \bibinfo{author}{\bibfnamefont{T.}~\bibnamefont{Jarlborg}}, \bibnamefont{and}
  \bibinfo{author}{\bibfnamefont{D.~N.} \bibnamefont{Timms}},
  \bibinfo{journal}{J. Phys.: Condensed Matter} \textbf{\bibinfo{volume}{10}},
  \bibinfo{pages}{2759} (\bibinfo{year}{1998}).

\bibitem[{\citenamefont{Timms et~al.}(1990)\citenamefont{Timms, Brahmia,
  Cooper, Collins, Hamouda, Laundy, Kilbourne, and Larger}}]{TB90}
\bibinfo{author}{\bibfnamefont{D.~N.} \bibnamefont{Timms}},
  \bibinfo{author}{\bibfnamefont{A.}~\bibnamefont{Brahmia}},
  \bibinfo{author}{\bibfnamefont{M.~J.} \bibnamefont{Cooper}},
  \bibinfo{author}{\bibfnamefont{S.~P.} \bibnamefont{Collins}},
  \bibinfo{author}{\bibfnamefont{S.}~\bibnamefont{Hamouda}},
  \bibinfo{author}{\bibfnamefont{D.}~\bibnamefont{Laundy}},
  \bibinfo{author}{\bibfnamefont{C.}~\bibnamefont{Kilbourne}},
  \bibnamefont{and} \bibinfo{author}{\bibfnamefont{M.-C.~S.}
  \bibnamefont{Larger}}, \bibinfo{journal}{J. Phys.: Condensed Matter}
  \textbf{\bibinfo{volume}{2}}, \bibinfo{pages}{3427} (\bibinfo{year}{1990}).

\bibitem[{\citenamefont{Eisenberger et~al.}(1972)\citenamefont{Eisenberger,
  Lam, Platzman, and Schmidt}}]{ei.la.72}
\bibinfo{author}{\bibfnamefont{P.}~\bibnamefont{Eisenberger}},
  \bibinfo{author}{\bibfnamefont{L.}~\bibnamefont{Lam}},
  \bibinfo{author}{\bibfnamefont{P.~M.} \bibnamefont{Platzman}},
  \bibnamefont{and} \bibinfo{author}{\bibfnamefont{P.}~\bibnamefont{Schmidt}},
  \bibinfo{journal}{Phys. Rev. B} \textbf{\bibinfo{volume}{6}},
  \bibinfo{pages}{3671} (\bibinfo{year}{1972}).

\bibitem[{\citenamefont{Lundqvist and Lyd\'en}(1971)}]{lu.ly.71}
\bibinfo{author}{\bibfnamefont{B.~I.} \bibnamefont{Lundqvist}}
  \bibnamefont{and} \bibinfo{author}{\bibfnamefont{C.}~\bibnamefont{Lyd\'en}},
  \bibinfo{journal}{Phys. Rev. B} \textbf{\bibinfo{volume}{4}},
  \bibinfo{pages}{3360} (\bibinfo{year}{1971}).

\bibitem[{\citenamefont{Bauer and Schneider}(1983)}]{BS83}
\bibinfo{author}{\bibfnamefont{G.~E.~W.} \bibnamefont{Bauer}} \bibnamefont{and}
  \bibinfo{author}{\bibfnamefont{J.~R.} \bibnamefont{Schneider}},
  \bibinfo{journal}{Z. Physik B} \textbf{\bibinfo{volume}{54}},
  \bibinfo{pages}{17} (\bibinfo{year}{1983}).

\bibitem[{\citenamefont{Bauer}(1984)}]{B84}
\bibinfo{author}{\bibfnamefont{G.~E.~W.} \bibnamefont{Bauer}},
  \bibinfo{journal}{Phys. Rev. B} \textbf{\bibinfo{volume}{30}},
  \bibinfo{pages}{1010} (\bibinfo{year}{1984}).

\bibitem[{\citenamefont{Bauer and Schneider}(1984)}]{BS84}
\bibinfo{author}{\bibfnamefont{G.~E.~W.} \bibnamefont{Bauer}} \bibnamefont{and}
  \bibinfo{author}{\bibfnamefont{J.~R.} \bibnamefont{Schneider}},
  \bibinfo{journal}{Phys. Rev. Lett.} \textbf{\bibinfo{volume}{52}},
  \bibinfo{pages}{2061} (\bibinfo{year}{1984}).

\bibitem[{\citenamefont{Bauer and Schneider}(1985)}]{BS85}
\bibinfo{author}{\bibfnamefont{G.~E.~W.} \bibnamefont{Bauer}} \bibnamefont{and}
  \bibinfo{author}{\bibfnamefont{J.~R.} \bibnamefont{Schneider}},
  \bibinfo{journal}{Phys. Rev. B} \textbf{\bibinfo{volume}{31}},
  \bibinfo{pages}{681} (\bibinfo{year}{1985}).

\bibitem[{\citenamefont{Benea et~al.}(2012)\citenamefont{Benea, Min\'ar,
  Chioncel, Mankovsky, and Ebert}}]{BMC+12}
\bibinfo{author}{\bibfnamefont{D.}~\bibnamefont{Benea}},
  \bibinfo{author}{\bibfnamefont{J.}~\bibnamefont{Min\'ar}},
  \bibinfo{author}{\bibfnamefont{L.}~\bibnamefont{Chioncel}},
  \bibinfo{author}{\bibfnamefont{S.}~\bibnamefont{Mankovsky}},
  \bibnamefont{and} \bibinfo{author}{\bibfnamefont{H.}~\bibnamefont{Ebert}},
  \bibinfo{journal}{Phys. Rev. B} \textbf{\bibinfo{volume}{85}},
  \bibinfo{pages}{085109} (\bibinfo{year}{2012}).

\bibitem[{\citenamefont{Chioncel et~al.}(2014)\citenamefont{Chioncel, Benea,
  Ebert, Di~Marco, and Min\'ar}}]{CBE+14}
\bibinfo{author}{\bibfnamefont{L.}~\bibnamefont{Chioncel}},
  \bibinfo{author}{\bibfnamefont{D.}~\bibnamefont{Benea}},
  \bibinfo{author}{\bibfnamefont{H.}~\bibnamefont{Ebert}},
  \bibinfo{author}{\bibfnamefont{I.}~\bibnamefont{Di~Marco}}, \bibnamefont{and}
  \bibinfo{author}{\bibfnamefont{J.}~\bibnamefont{Min\'ar}},
  \bibinfo{journal}{Phys. Rev. B} \textbf{\bibinfo{volume}{85}},
  \bibinfo{pages}{085109} (\bibinfo{year}{2014}).

\bibitem[{\citenamefont{Ebert et~al.}(2011)\citenamefont{Ebert, K\"odderitzsch,
  and Min\'{a}r}}]{EKM11}
\bibinfo{author}{\bibfnamefont{H.}~\bibnamefont{Ebert}},
  \bibinfo{author}{\bibfnamefont{D.}~\bibnamefont{K\"odderitzsch}},
  \bibnamefont{and}
  \bibinfo{author}{\bibfnamefont{J.}~\bibnamefont{Min\'{a}r}},
  \bibinfo{journal}{Rep. Prog. Phys.} \textbf{\bibinfo{volume}{74}},
  \bibinfo{pages}{096501} (\bibinfo{year}{2011}).

\bibitem[{\citenamefont{Vosko et~al.}(1980)\citenamefont{Vosko, Wilk, and
  Nusair}}]{VWN80}
\bibinfo{author}{\bibfnamefont{S.~H.} \bibnamefont{Vosko}},
  \bibinfo{author}{\bibfnamefont{L.}~\bibnamefont{Wilk}}, \bibnamefont{and}
  \bibinfo{author}{\bibfnamefont{M.}~\bibnamefont{Nusair}},
  \bibinfo{journal}{Can. J. Phys.} \textbf{\bibinfo{volume}{58}},
  \bibinfo{pages}{1200} (\bibinfo{year}{1980}).

\bibitem[{\citenamefont{Monkhorst and Pack}(1976)}]{MP76}
\bibinfo{author}{\bibfnamefont{H.~J.} \bibnamefont{Monkhorst}}
  \bibnamefont{and} \bibinfo{author}{\bibfnamefont{J.~D.} \bibnamefont{Pack}},
  \bibinfo{journal}{Phys. Rev. B} \textbf{\bibinfo{volume}{13}},
  \bibinfo{pages}{5188} (\bibinfo{year}{1976}).

\bibitem[{\citenamefont{Katsnelson and Lichtenstein}(2002)}]{KL02}
\bibinfo{author}{\bibfnamefont{M.~I.} \bibnamefont{Katsnelson}}
  \bibnamefont{and} \bibinfo{author}{\bibfnamefont{A.~I.}
  \bibnamefont{Lichtenstein}}, \bibinfo{journal}{Eur. Phys. J. B}
  (\bibinfo{year}{2002}).

\bibitem[{\citenamefont{Pourovskii et~al.}(2005)\citenamefont{Pourovskii,
  Katsnelson, and Lichtenstein}}]{PKL05}
\bibinfo{author}{\bibfnamefont{L.~V.} \bibnamefont{Pourovskii}},
  \bibinfo{author}{\bibfnamefont{M.~I.} \bibnamefont{Katsnelson}},
  \bibnamefont{and} \bibinfo{author}{\bibfnamefont{A.~I.}
  \bibnamefont{Lichtenstein}}, \bibinfo{journal}{Phys. Rev. B}
  (\bibinfo{year}{2005}).

\bibitem[{\citenamefont{Marco et~al.}(2009)\citenamefont{Marco, Min\'ar,
  Chadov, Katsnelson, Ebert, and Lichtenstein}}]{MM09}
\bibinfo{author}{\bibfnamefont{I.~D.} \bibnamefont{Marco}},
  \bibinfo{author}{\bibfnamefont{J.}~\bibnamefont{Min\'ar}},
  \bibinfo{author}{\bibfnamefont{S.}~\bibnamefont{Chadov}},
  \bibinfo{author}{\bibfnamefont{M.~I.} \bibnamefont{Katsnelson}},
  \bibinfo{author}{\bibfnamefont{H.}~\bibnamefont{Ebert}}, \bibnamefont{and}
  \bibinfo{author}{\bibfnamefont{A.~I.} \bibnamefont{Lichtenstein}},
  \bibinfo{journal}{Phys. Rev. B} \textbf{\bibinfo{volume}{79}},
  \bibinfo{pages}{115111} (\bibinfo{year}{2009}).

\bibitem[{\citenamefont{Min\'ar}(2011)}]{MI11}
\bibinfo{author}{\bibfnamefont{J.}~\bibnamefont{Min\'ar}}, \bibinfo{journal}{J.
  Phys.: Condensed Matter} \textbf{\bibinfo{volume}{23}},
  \bibinfo{pages}{253201} (\bibinfo{year}{2011}).

\bibitem[{\citenamefont{Aryasetiawan et~al.}(2004)\citenamefont{Aryasetiawan,
  Imada, Georges, Kotliar, Biermann, and Lichtenstein}}]{AIG+04}
\bibinfo{author}{\bibfnamefont{F.}~\bibnamefont{Aryasetiawan}},
  \bibinfo{author}{\bibfnamefont{M.}~\bibnamefont{Imada}},
  \bibinfo{author}{\bibfnamefont{A.}~\bibnamefont{Georges}},
  \bibinfo{author}{\bibfnamefont{G.}~\bibnamefont{Kotliar}},
  \bibinfo{author}{\bibfnamefont{S.}~\bibnamefont{Biermann}}, \bibnamefont{and}
  \bibinfo{author}{\bibfnamefont{A.~I.} \bibnamefont{Lichtenstein}},
  \bibinfo{journal}{Phys. Rev. B} \textbf{\bibinfo{volume}{70}},
  \bibinfo{pages}{195104} (\bibinfo{year}{2004}).

\bibitem[{\citenamefont{Szotek et~al.}(1984)\citenamefont{Szotek, Gyorffy,
  Stocks, and Temmerman}}]{SGST84}
\bibinfo{author}{\bibfnamefont{Z.}~\bibnamefont{Szotek}},
  \bibinfo{author}{\bibfnamefont{B.~L.} \bibnamefont{Gyorffy}},
  \bibinfo{author}{\bibfnamefont{G.~M.} \bibnamefont{Stocks}},
  \bibnamefont{and} \bibinfo{author}{\bibfnamefont{W.~M.}
  \bibnamefont{Temmerman}}, \bibinfo{journal}{J. Phys. F: Met. Phys.}
  \textbf{\bibinfo{volume}{14}}, \bibinfo{pages}{2571} (\bibinfo{year}{1984}).

\bibitem[{\citenamefont{Benea et~al.}(2006)\citenamefont{Benea, Mankovsky, and
  Ebert}}]{BME06}
\bibinfo{author}{\bibfnamefont{D.}~\bibnamefont{Benea}},
  \bibinfo{author}{\bibfnamefont{S.}~\bibnamefont{Mankovsky}},
  \bibnamefont{and} \bibinfo{author}{\bibfnamefont{H.}~\bibnamefont{Ebert}},
  \bibinfo{journal}{Phys. Rev. B} \textbf{\bibinfo{volume}{73}},
  \bibinfo{pages}{094411} (\bibinfo{year}{2006}).

\bibitem[{\citenamefont{Benea}(2004)}]{DB04}
\bibinfo{author}{\bibfnamefont{D.}~\bibnamefont{Benea}}, Ph.D. thesis,
  \bibinfo{school}{LMU M\"unchen} (\bibinfo{year}{2004}).

\bibitem[{\citenamefont{Baruah et~al.}(2000)\citenamefont{Baruah, Zope, and
  Kshirsagar}}]{BZK00}
\bibinfo{author}{\bibfnamefont{T.}~\bibnamefont{Baruah}},
  \bibinfo{author}{\bibfnamefont{R.~R.} \bibnamefont{Zope}}, \bibnamefont{and}
  \bibinfo{author}{\bibfnamefont{A.}~\bibnamefont{Kshirsagar}},
  \bibinfo{journal}{Phys. Rev. B} \textbf{\bibinfo{volume}{62}},
  \bibinfo{pages}{16435} (\bibinfo{year}{2000}).

\bibitem[{\citenamefont{Kubo}(2004)}]{K04}
\bibinfo{author}{\bibfnamefont{Y.}~\bibnamefont{Kubo}}, \bibinfo{journal}{J.
  Phys. Chem. Solids} \textbf{\bibinfo{volume}{65}}, \bibinfo{pages}{2077}
  (\bibinfo{year}{2004}).

\bibitem[{\citenamefont{Kakutani et~al.}(2003)\citenamefont{Kakutani, Kubo,
  Koizumi, Sakai, Ahuja, and Sarma}}]{KKK+03}
\bibinfo{author}{\bibfnamefont{Y.}~\bibnamefont{Kakutani}},
  \bibinfo{author}{\bibfnamefont{Y.}~\bibnamefont{Kubo}},
  \bibinfo{author}{\bibfnamefont{A.}~\bibnamefont{Koizumi}},
  \bibinfo{author}{\bibfnamefont{N.}~\bibnamefont{Sakai}},
  \bibinfo{author}{\bibfnamefont{B.~L.} \bibnamefont{Ahuja}}, \bibnamefont{and}
  \bibinfo{author}{\bibfnamefont{B.~K.} \bibnamefont{Sarma}},
  \bibinfo{journal}{J. Phys. Soc. Japan} \textbf{\bibinfo{volume}{72}},
  \bibinfo{pages}{599} (\bibinfo{year}{2003}).

\bibitem[{\citenamefont{Majundar}(1965)}]{M65}
\bibinfo{author}{\bibfnamefont{C.~K.} \bibnamefont{Majundar}},
  \bibinfo{journal}{Phys. Rev.} \textbf{\bibinfo{volume}{140}},
  \bibinfo{pages}{A227} (\bibinfo{year}{1965}).

\bibitem[{\citenamefont{Eisenberger and Reed}(1974)}]{ei.re.74}
\bibinfo{author}{\bibfnamefont{P.}~\bibnamefont{Eisenberger}} \bibnamefont{and}
  \bibinfo{author}{\bibfnamefont{W.~A.} \bibnamefont{Reed}},
  \bibinfo{journal}{Phys. Rev. B} \textbf{\bibinfo{volume}{9}},
  \bibinfo{pages}{3242} (\bibinfo{year}{1974}).

\bibitem[{\citenamefont{Wang and Callaway}(1975)}]{wa.ca.75}
\bibinfo{author}{\bibfnamefont{C.~S.} \bibnamefont{Wang}} \bibnamefont{and}
  \bibinfo{author}{\bibfnamefont{J.}~\bibnamefont{Callaway}},
  \bibinfo{journal}{Phys. Rev. B} \textbf{\bibinfo{volume}{11}},
  \bibinfo{pages}{2417} (\bibinfo{year}{1975}).

\bibitem[{\citenamefont{Anastassopoulos
  et~al.}(1990)\citenamefont{Anastassopoulos, Priftis, Papanicolau, Bacalis,
  and Papaconstantopoulos}}]{AP90}
\bibinfo{author}{\bibfnamefont{D.~L.} \bibnamefont{Anastassopoulos}},
  \bibinfo{author}{\bibfnamefont{G.~D.} \bibnamefont{Priftis}},
  \bibinfo{author}{\bibfnamefont{N.~I.} \bibnamefont{Papanicolau}},
  \bibinfo{author}{\bibfnamefont{N.~C.} \bibnamefont{Bacalis}},
  \bibnamefont{and} \bibinfo{author}{\bibfnamefont{D.~A.}
  \bibnamefont{Papaconstantopoulos}}, \bibinfo{journal}{J. Phys.: Condensed
  Matter} \textbf{\bibinfo{volume}{3}}, \bibinfo{pages}{1099}
  (\bibinfo{year}{1990}).

\bibitem[{\citenamefont{Rollason et~al.}(1987)\citenamefont{Rollason,
  Schneider, Laundy, Holt, and Cooper}}]{RS87}
\bibinfo{author}{\bibfnamefont{A.~J.} \bibnamefont{Rollason}},
  \bibinfo{author}{\bibfnamefont{J.~R.} \bibnamefont{Schneider}},
  \bibinfo{author}{\bibfnamefont{D.~S.} \bibnamefont{Laundy}},
  \bibinfo{author}{\bibfnamefont{R.~S.} \bibnamefont{Holt}}, \bibnamefont{and}
  \bibinfo{author}{\bibfnamefont{M.~J.} \bibnamefont{Cooper}},
  \bibinfo{journal}{J. Phys. F: Met. Phys.} \textbf{\bibinfo{volume}{17}},
  \bibinfo{pages}{1105} (\bibinfo{year}{1987}).

\bibitem[{\citenamefont{Strange}(1998)}]{T3}
\bibinfo{author}{\bibfnamefont{P.}~\bibnamefont{Strange}},
  \emph{\bibinfo{title}{Relativistic Quantum Mechanics}}
  (\bibinfo{publisher}{Cambridge: University Press}, \bibinfo{year}{1998}).

\bibitem[{\citenamefont{Ebert et~al.}(1997)\citenamefont{Ebert, Freyer, and
  Deng}}]{eb.fr.97}
\bibinfo{author}{\bibfnamefont{H.}~\bibnamefont{Ebert}},
  \bibinfo{author}{\bibfnamefont{H.}~\bibnamefont{Freyer}}, \bibnamefont{and}
  \bibinfo{author}{\bibfnamefont{M.}~\bibnamefont{Deng}},
  \bibinfo{journal}{Phys. Rev. B} \textbf{\bibinfo{volume}{56}},
  \bibinfo{pages}{9454} (\bibinfo{year}{1997}).

\bibitem[{\citenamefont{Ebert}(2000)}]{E00}
\bibinfo{author}{\bibfnamefont{H.}~\bibnamefont{Ebert}}, in
  \emph{\bibinfo{booktitle}{Electronic Structure and Physical Properties of
  Solids}}, edited by
  \bibinfo{editor}{\bibfnamefont{H.}~\bibnamefont{Dreyss\'{e}}}
  (\bibinfo{publisher}{Springer, Berlin}, \bibinfo{year}{2000}), vol.
  \bibinfo{volume}{535}, p. \bibinfo{pages}{191}.

\bibitem[{\citenamefont{Dirac}(1928)}]{D28}
\bibinfo{author}{\bibfnamefont{P.~A.~M.} \bibnamefont{Dirac}},
  \bibinfo{journal}{Proc. Roy. Soc. (London) A} \textbf{\bibinfo{volume}{118}},
  \bibinfo{pages}{351} (\bibinfo{year}{1928}).

\bibitem[{\citenamefont{Foldy and Wouthuysen}(1950)}]{FW50}
\bibinfo{author}{\bibfnamefont{L.~L.} \bibnamefont{Foldy}} \bibnamefont{and}
  \bibinfo{author}{\bibfnamefont{S.~A.} \bibnamefont{Wouthuysen}},
  \bibinfo{journal}{Phys. Rev.} \textbf{\bibinfo{volume}{78}},
  \bibinfo{pages}{29} (\bibinfo{year}{1950}).

\bibitem[{\citenamefont{Kotliar et~al.}(2006)\citenamefont{Kotliar, Savrasov,
  Haule, Oudovenko, Parcollet, and Marianetti}}]{KS06}
\bibinfo{author}{\bibfnamefont{G.}~\bibnamefont{Kotliar}},
  \bibinfo{author}{\bibfnamefont{S.~Y.} \bibnamefont{Savrasov}},
  \bibinfo{author}{\bibfnamefont{K.}~\bibnamefont{Haule}},
  \bibinfo{author}{\bibfnamefont{V.~S.} \bibnamefont{Oudovenko}},
  \bibinfo{author}{\bibfnamefont{O.}~\bibnamefont{Parcollet}},
  \bibnamefont{and} \bibinfo{author}{\bibfnamefont{C.~A.}
  \bibnamefont{Marianetti}}, \bibinfo{journal}{Rev. Mod. Phys.}
  \textbf{\bibinfo{volume}{78}}, \bibinfo{pages}{865} (\bibinfo{year}{2006}).

\bibitem[{\citenamefont{Cooper}(1985)}]{C85}
\bibinfo{author}{\bibfnamefont{M.~J.} \bibnamefont{Cooper}},
  \bibinfo{journal}{Rep. Prog. Phys.} \textbf{\bibinfo{volume}{48}},
  \bibinfo{pages}{415} (\bibinfo{year}{1985}).

\end{thebibliography}

\end{document}